\documentclass[a4paper]{aa}

\usepackage{graphicx,natbib}
\usepackage{txfonts}
\usepackage{supertabular}

\newcommand{\Bz}{$\langle B_\mathrm{z} \rangle$}

\begin{document}

   \title{The search for magnetic fields in mercury-manganese stars\thanks{Based on observations collected at the European Southern Observatory, Chile (ESO programs 083.D-1000, 084.D-0338, 085.D-0296).}}

   \subtitle{}

   \author{V.~Makaganiuk\inst{1}
      \and O.~Kochukhov\inst{1}
      \and N.~Piskunov\inst{1}
      \and S.~V.~Jeffers\inst{2}
      \and C.~M.~Johns-Krull\inst{4}
      \and C.~U.~Keller\inst{2}
      \and M.~Rodenhuis\inst{2}
      \and F.~Snik\inst{2}
      \and H.~C.~Stempels\inst{1}
      \and J.~A.~Valenti\inst{3}}

   \institute{Department Physics and Astronomy, Uppsala University, Box 516, 751 20 Uppsala, Sweden
   \and
Sterrekundig Instituut, Universiteit Utrecht, P.O. Box 80000, NL-3508 TA Utrecht, The Netherlands
   \and
Space Telescope Science Institute, 3700 San Martin Dr, Baltimore MD 21211, USA
   \and
Department of Physics and Astronomy, Rice University, 6100 Main Street, Houston, TX 77005, USA}

   \date{Received 31 August 2010 / Accepted 18 October 2010}

  \abstract
   {A subclass of the upper main sequence chemically peculiar stars, mercury-manganese (HgMn) stars were traditionally considered to be non-magnetic, showing no evidence for variability in spectral line profiles. However, recent discoveries of chemical inhomogeneities on their surfaces call for revision of this question. In particular, spectroscopic time-series of AR~Aur, $\alpha$~And, and five other HgMn stars indicate the presence of chemical spots. At the same time, no signatures of global magnetic fields were detected.}
   {In order to understand the physical mechanism that causes the formation of chemical spots in HgMn stars and to gain insight into the potential magnetic field properties at their surfaces, we performed a highly sensitive search for magnetic fields on a large set of HgMn stars.}
   {With the aid of a new polarimeter attached to the HARPS spectrometer at the ESO 3.6m-telescope, we obtained high-quality circular polarization spectra of 41 single and double HgMn stars. Using a multi-line analysis technique on each star, we co-added information from hundreds of spectral lines resulting in significantly greater sensitivity to the presence of magnetic fields, including very weak fields.}
   {For the 47 individual objects studied, including 6 components of SB2 systems, we do not detect any magnetic fields at greater than the 3$\sigma$ level. The lack of detection in the circular polarization profiles indicates that if strong fields are present on these stars, they must have complex surface topologies. For simple global fields, our detection limits imply upper limits to the fields present of 2--10 Gauss in the best cases.}
   {We conclude that HgMn stars lack large-scale magnetic fields, typical for spotted magnetic Ap stars, sufficient to form and sustain the chemical spots observed on HgMn stars. Our study confirms that in addition to magnetically altered atomic diffusion, there exists another differentiation mechanism operating in the atmospheres of late-B main sequence stars which can compositional inhomogeneities on their surfaces.}


   \maketitle


\section{Introduction}
Mercury-manganese (HgMn) stars form a subclass of the upper main sequence chemically peculiar (CP) stars, showing a notable overabundance of Hg, Mn, Y, Sr and other, mostly heavy, chemical elements with respect to the solar chemical composition. HgMn stars are frequently found in binaries and lie on the H-R diagram between the early-A and late-B spectral types, which corresponds to $T_\mathrm{eff}$\,=\,9500--16000~K \citep{Dworetsky:1993}.

While most other, normal stars in this spectral range are rapid rotators with  $v_\mathrm{e}\sin{i}=200-300$~km\,s$^{-1}$ \citep{Abt:1995}, HgMn stars are typically slow rotators, making them ideal late-B targets for detailed abundance analyses.  The HgMn stars were for a long time considered to have chemically homogeneous atmospheres, with no obvious vertical chemical abundance stratification or horizontal concentrations of some elements into surface spots of high elemental abundance such as those in magnetic Ap stars of similar temperatures \citep{Kochukhov:2004}. This view was historically supported by the lack of a definite detection of line profile variability in any HgMn star. However, recently \citet{Adelman:2002} reported variability in the spectral line of \ion{Hg}{ii} $\lambda$ 3984~\AA\ based on high-resolution and high signal-to-noise (S/N) ratio spectroscopic data obtained for the brightest HgMn star: $\alpha$ And. They attributed the variability in the \ion{Hg}{ii} line to horizontal inhomogeneities of the mercury abundance across the stellar surface and reconstructed a surface map of Hg with the help of Doppler imaging.

Later, \citet{Kochukhov:2005} found two other HgMn stars (HR~1185 and HR~8723) possessing spotted structure in their abundance of Hg. The fourth star, the eclipsing HgMn binary AR~Aur studied by \citet{Hubrig:2006}, was reported to exhibit line profile variability of Hg and also such chemical elements as Y, Zr, Pt, and Sr. These results were confirmed by the independent study of \citet{Folsom:2010}.

An additional unexpected discovery was made by \citet{Kochukhov:2007} for the HgMn star $\alpha$ And. From the analysis of the \ion{Hg}{ii} 3984~\AA\ line observed over a timespan of 7 years they inferred surface maps which display morphological changes of spots on a time scale of only a few years. In comparison to well-studied magnetic CP stars, this behavior of the surface abundance distribution is quite surprising because all other early-type spotted stars show spot configurations that are stable over tens of years \citep[e.g.,][]{Adelman:2001}.

Finally, \citet{Briquet:2010} investigated time-series spectra of three more HgMn stars and concluded that all of them display variability in their line profiles. The star HD~11753 showed variability in Y, Sr and Ti; while HD~53244 and HD~221507 both show variable profiles of their Mn and Hg lines.  In addition, HD~221507 displays variations in its Y lines.
In summary, these recent studies show there are at least 7 spotted HgMn stars and possibly many more. It is still not clear how typical this behavior is for this class of CP stars.  Furthermore, it is not clear what physical processes are responsible for the formation of these structures on the stellar surface.

It is generally believed that a strong magnetic field is a necessary ingredient for creating inhomogeneities in the stellar atmosphere, leading, for example, to the formation of temperature or chemical spots \citep[e.g.,][]{Jeffers:2008, Luftinger:2010}. Within this framework, a number of spectropolarimetric observations have been carried out on HgMn stars in order to confirm or disprove the presence of magnetic fields on these stars which could be responsible for the formation of chemical spots. The first systematic attempt to detect magnetic signatures in high-resolution circularly polarized spectra of HgMn stars was made by \citet{Shorlin:2002}. This study included a sample of 10 stars which were analysed at a resolution of $R\,=\,35\,000$, yielding upper limits for the magnetic field strength of 29 to 100~G. To date, this is the only extensive magnetic survey of HgMn stars. Several other studies searched for magnetic fields in individual spotted HgMn stars. For instance, \citet{Wade:2006} searched for magnetic a field in $\alpha$~And. They analysed magnetic field measurements obtained with three different polarimeters and concluded that the star has no longitudinal magnetic field stronger than about 6--19~G. \citet{Folsom:2010} performed similar analysis for AR~Aur, once again finding no field stronger than 20--40~G. Finally, \citet{Auriere:2010} presented an analysis of a small sample of three bright, sharp-lined HgMn stars, reporting no magnetic field detections at the level of 1--3~G.

Previous magnetic field studies of HgMn stars suffered from several fundamental limitations. First, with the exception of the four stars studied by \citet{Wade:2006} and \citet{Auriere:2010}, previous work was not particularly precise, possibly missing weak magnetic fields which are still capable of causing chemical spot formation. Second, they included a small number of HgMn stars, often with a strong emphasis on the most slowly rotating ones.  Thus, the class as a whole has not been well surveyed.

To overcome drawbacks associated with previous magnetic field studies on these stars, and to provide new insights into the spot formation physics of late-B stars, we have initiated a spectropolarimetric survey of a large sample of HgMn stars. We investigate HgMn stars with a broad range of atmospheric parameters and rotational velocities. Most of these objects have not been studied before with the high-resolution spectropolarimetry. Taking advantage of a new polarimeter, HARPSpol, attached to the HARPS instrument at the ESO 3.6-m telescope in La~Silla, Chile, we were able to push the limits of magnetic field detection in early-type stars down to remarkably low levels.

In Sect.~\ref{targets} we describe our target selection procedure. Sect.~\ref{observations} discusses our spectropolarimetric observations. Sect.~\ref{reduction} outlines the principles of the reduction of the spectropolarimetric data from HARPSpol. We discuss our multi-line magnetic field detection technique in Sect.~\ref{lsd} and present the results of our magnetic field measurements in Sect.~\ref{mf}. Sect.~\ref{disc} summarizes our results and discusses them in the context of recent studies of HgMn stars.

\section{Observations}
\subsection{Target selection}
\label{targets}

To compile our target list we used the catalogue of Ap, HgMn, and Am stars published by \citet{Renson:2009}. In order to optimize our observations for the detection of weak magnetic fields and to cover a statistically large sample of HgMn stars, we put several constraints on the stellar parameters. All objects with $v_\mathrm{e}\sin{i}\ga$\,70~km\,s$^{-1}$ were excluded because magnetic field measurements are considerably less accurate for rapid rotators in comparison to the moderately and, especially, slowly rotating stars.
We also preferentially selected brighter stars. Aiming to reach a signal-to-noise ratio ($S/N$) of $\approx$\,300, we selected only targets brighter than $m_\mathrm{V}=7$.  A total of 45 HgMn stars were found that satisfied these criteria. Our three observing runs covered 41 of them. Among these stars 6 are spectroscopic binary systems, allowing magnetic field measurements on both components.

\begin{table}[!t]
   \caption{The list of HgMn stars included in our survey.}
   \label{tab1}
\centering
\begin{tabular}{l l l c c}
\hline\hline
HD number & HR number & $m_\mathrm{V}$ & $T_\mathrm{eff}$ (K) & Binarity \\
\hline
HD~1909   & HR~89   & $6.56$ & 12406 &  SB2 \\
HD~11753  & HR~558  & $5.11$ & 10476 &  SB1 \\
HD~27376  & HR~1347 & $3.55$ & 12849 &  SB2 \\
HD~28217  & HR~1402 & $5.87$ & 13906 &  SB1 \\
HD~29589  & HR~1484 & $5.45$ & 14763 &      \\
HD~32964  & HR~1657 & $5.10$ & 11119 &  SB2 \\
HD~33647  & HR~1690 & $6.67$ & 12440 &  SB2 \\
HD~33904  & HR~1702 & $3.28$ & 12759 &      \\
HD~34880  & HR~1759 & $6.41$ & 13269 &      \\
HD~35548  & HR~1800 & $6.56$ & 11164 &  SB2 \\
HD~36881  & HR~1883 & $5.63$ & 10903 &  SB1 \\
HD~42657  & HR~2202 & $6.20$ & 12842 &      \\
HD~53244  & HR~2657 & $4.10$ & 13596 &      \\
HD~53929  & HR~2676 & $6.09$ & 13908 &      \\
HD~63975  & HR~3059 & $5.13$ & 13458 &      \\
HD~65950  &         & $6.87$ & 12710 &      \\
HD~68099  & HR~3201 & $6.08$ & 12997 &      \\
HD~70235  & HR~3273 & $6.43$ & 12329 &      \\
HD~71066  & HR~3302 & $5.62$ & 12131 &      \\
HD~71833  & HR~3345 & $6.67$ & 12985 &      \\
HD~72208  & HR~3361 & $6.83$ & 11141 &  SB2 \\
HD~75333  & HR~3500 & $5.31$ & 12248 &      \\
HD~78316  & HR~3623 & $5.24$ & 13639 &  SB2 \\
HD~90264  & HR~4089 & $4.95$ & 15121 &  SB2 \\
HD~101189 & HR~4487 & $5.14$ & 11148 &      \\
HD~106625 & HR~4662 & $2.59$ & 12002 &  SB1 \\
HD~110073 & HR~4817 & $4.63$ & 12876 &      \\
HD~141556 & HR~5883 & $3.96$ & 10684 &  SB2 \\
HD~149121 & HR~6158 & $5.62$ & 11021 &  SB1 \\
HD~158704 & HR~6520 & $6.06$ & 13378 &  SB2 \\
HD~165493 & HR~6759 & $6.15$ & 14375 &  SB2 \\
HD~175640 & HR~7143 & $6.20$ & 12075 &      \\
HD~178065 & HR~7245 & $6.56$ & 12348 &  SB1 \\
HD~179761 & HR~7287 & $5.14$ & 13010 &      \\
HD~186122 & HR~7493 & $6.33$ & 12901 &      \\
HD~191110 & HR~7694 & $6.18$ & 12107 &  SB2 \\
HD~193452 & HR~7775 & $6.10$ & 10881 &  SB1 \\
HD~194783 & HR~7817 & $6.08$ & 13803 &      \\
HD~202671 & HR~8137 & $5.39$ & 13696 &      \\
HD~211838 & HR~8512 & $5.35$ & 12593 &  SB1 \\
HD~221507 & HR~8937 & $4.37$ & 12476 &      \\
\hline
\end{tabular}
\end{table}

In the Table~\ref{tab1} we present the list of observed stars, giving their HD and HR numbers, visual magnitude, effective temperature and binary status. The effective temperature of all objects was estimated using the \citet{MD85} calibration of the Str\"omgren photometric parameters as implemented in the TempLogG code \citep{Kupka:2001}. The  Str\"omgren photometry was obtained from the Simbad\footnote{http://simbad.u-strasbg.fr/simbad/} database. Information about binarity was extracted from the catalogue of \citet{SB9} and complemented by the results of several recent studies of individual binary systems. For a few stars we used our LSD line profiles as described below to detect binarity.

\subsection{Spectropolarimetric observations}
\label{observations}

Our observations were obtained using the newly built polarimeter HARPSpol \citep{HARPSpol} attached to the HARPS spectrometer \citep{HARPS} at the ESO 3.6-m telescope. With a resolving power of $R$\,=\,115\,000, this instrument is the highest resolution spectropolarimeter available to the astronomical community.
The compact optical design of the polarimeter allows it to be mounted before the fiber entrance at the Cassegrain focus of the telescope. This positioning of the polarimeter minimizes the instrumental polarization which is usually caused by the oblique reflections in the light path. HARPSpol consists of two independent polarimeters for circular and linear polarization measurements. Thus, it is possible to make observations in all four Stokes parameters. Each polarimeter consists of a superachromatic retarder plate (a half-wave or a quarter-wave plate) and a beam splitter based on a Foster prism. This design makes the polarimeter fully achromatic. The two polarized light spectra are recorded simultaneously through the two HARPS entrance fibers.

Our observations were carried out between May--June of 2009, January of 2010, and April--May of 2010.  The date each of our target stars was observed is recorded in Table 2.  In total, we have observed HgMn targets on 26 nights, which were shared with other spectropolarimetric programs. We obtained high-resolution spectra with a typical $S/N$ of 150--400. The spectra were recorded by a mosaic of two 2K~$\times$~4K CCDs, providing 45 polarimetric echelle orders on the blue CCD and 26 on the red one. For each night we have acquired a standard set of calibration images: 20 bias exposures, 20 flat fields and 2 ThAr frames. Flat field and ThAr images were exposed with the polarimeter in the circular polarization mode.

All HgMn stars were observed in circular polarization, covering the wavelength range of 3780--6913~\AA\ with a small gap at 5259--5337~\AA. Each observation of an individual star was split into four sub-exposures corresponding to differing positions of the quarter-wave plate: 45$\degr$, 135$\degr$, 225$\degr$ and 315$\degr$ relative to the optical axis of the beam-splitter. The length of individual sub-exposures ranged between 160 and 350~s. In a few cases stars were observed with only 2 sub-exposures.
The majority of our targets were observed only once; however, 10 stars, mainly spectroscopic binaries, were observed several times. For HD~11753 and HD~32964 we obtained complete coverage of the rotational and orbital periods respectively. Analysis of these HgMn stars will be presented in future publications.

\subsection{Reduction of the data}
\label{reduction}

The reduction of our HARPSpol observations was performed with the REDUCE package of \citet{reduce}. This set of IDL routines applies a standard sequence of reduction and calibration procedures to the cross-dispersed echelle spectra. Bias images are averaged and subtracted from the average flat and science images. Spectral orders are located using the average flat field with the help of a cluster analysis method. The flat field is normalized and is used to correct the pixel-to-pixel sensitivity variations in the science images. After the removal of scattered light, the science spectra are extracted using the optimal extraction algorithm described by \citet{reduce}.

The intrinsic long-term stability of the HARPS spectrometer is of the order of 1~m\,s$^{-1}$ \citep{HARPS}. We do not expect the polarimeter to adversely affect this stability because it does not significantly modify the seeing-limited illumination of the HARPS fibers \citep{HARPSpol}. Since we do not require an extremely high velocity accuracy for our observations, it was sufficient to use only one nightly ThAr spectrum for the wavelength calibration. Using $\approx$\,700--900 ThAr lines in a 2-D wavelength calibration routine, we obtained an internal wavelength calibration accuracy of 18--21~m~s$^{-1}$. In the final step before the calculation of the Stokes parameter spectra we performed continuum normalization, dividing each spectrum by a smooth, slowly varying function. This function was obtained by fitting the upper envelope of the blaze-corrected merged spectrum.

For the calculation of the circular polarization we use the ratio method described by \citet{Bagnulo:2009}. The method reduces the spurious polarization effects by an appropriate combination of the two physical beams recorded for four different retarder plate positions. Along with the circular polarization we also derive a diagnostic null spectrum. It is obtained by combining individual sub-exposures destructively, thus canceling the stellar polarization signal and showing the residual instrumental polarization remaining after application of the ratio method. Below we use the null spectrum in the same analysis steps as the Stokes~$V$ spectrum, thus providing a realistic estimate of possible errors.

Using the eight spectra in the four individual sub-exposures of each star provides a convenient means to detect and remove cosmic ray hits that otherwise seriously distort the final Stokes spectra. Affected pixels are identified by their large deviation from the median and are substituted by the latter. Since our exposure times were relatively short, only 1--2 pixels required correction in each echelle order.

\section{Data analysis and results}
\subsection{LSD analysis}
\label{lsd}

\begin{figure*}[!t]
   \centering
   {\resizebox{\hsize}{!}{\rotatebox{90}{\includegraphics{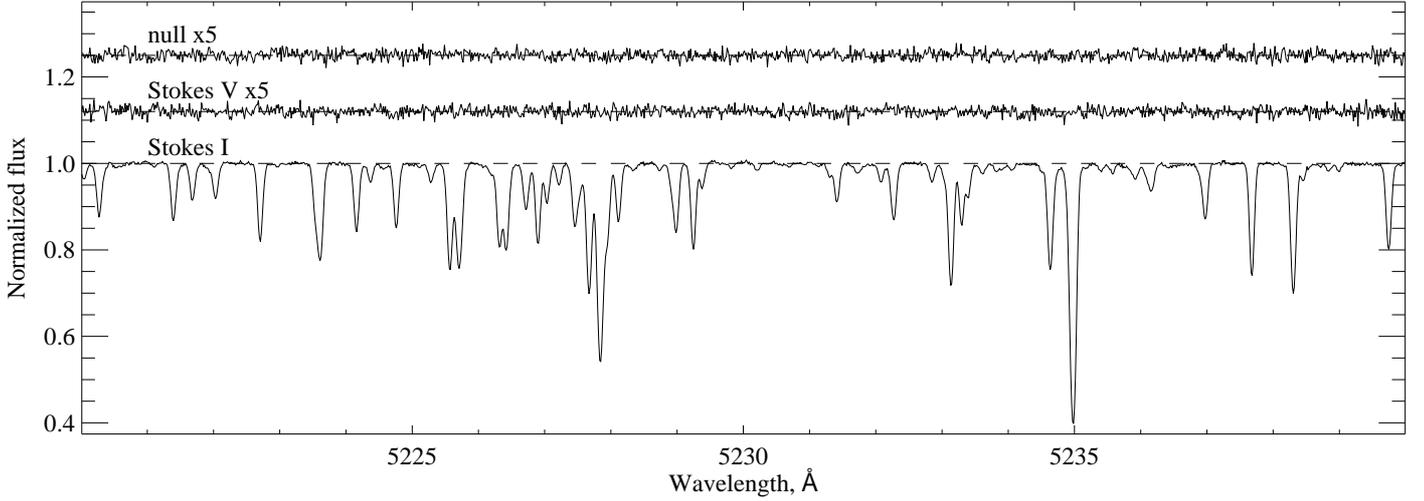}}}}
   \caption{The HARPSpol spectra of HD~71066 in the wavelength region of 5220--5240~\AA. Spectra are displayed from bottom to top as follows: Stokes~$I$, Stokes~$V$ and the null spectrum. Note, that for displaying purposes Stokes $V$ and the null spectrum are shifted upwards by 1.12 and 1.25, respectively, and are multiplied by 5.}
   \label{spec}
\end{figure*}

A major difficulty  when searching for weak (below 100~G) magnetic fields in stars is the weakness of expected polarimetric signal. As one can see from Fig.~\ref{spec}, it is difficult to conclude anything about the presence or absence of the magnetic signal in individual spectral lines even with our best Stokes~$V$ spectra.  To alleviate this problem, a line addition technique, called Least-Squares Deconvolution (hereafter LSD) developed by \citet{Donati:1997}, is commonly used to detect weak stellar magnetic fields. This technique has proven to be a very effective tool, allowing extraction of high-precision Stokes~$I$ and $V$ profiles by combining the information from all available metal lines in the spectrum. The main assumption of LSD is that all spectral lines are identical in shape and can be represented by a scaled average profile. Using this model one can reconstruct the average profile from an observed intensity or polarization spectrum, obtaining an increase of the signal-to-noise ratio by a factor of up to 30--40. Recent publications have demonstrated that with the help of the LSD technique, it is possible to measure magnetic fields weaker than 1~G \citep{Auriere:2009, Lignieres:2009}.

The LSD analysis requires prior knowledge of the line positions and strengths. We have compiled a number of line lists using VALD \citep{VALD} together with a 1-D, LTE stellar atmosphere models with effective temperatures ranging from 10000~K to 18000~K in 500~K steps. The atmosphere models were calculated with the LLmodels code \citep{LLmodels}, assuming solar abundances. When doing the spectrum synthesis to estimate central line depths in order to select lines to include in our line lists, the abundances of some chemical elements were changed with respect to the Sun to mimic the abundance pattern typical of an HgMn star, e.g. [Fe]\,=\,+0.2, [Cr]\,=\,+0.5, [Mg]\,=\,$-0.2$, [Ti]\,=\,+0.5.

The effective temperature of each star, derived from the Str\"omgren photometric colors, was rounded to the nearest value in the model grid. A separate line mask was then created for each stare. In order to exclude very weak lines from the calculation of LSD profiles, we required that the central depth of each line be $\geq 10$\% of the local continuum in order for the line to be included in the list.  For the coolest star in our sample ($T_\mathrm{eff}\approx10500$~K) this gave us 753 lines for the calculation of the LSD profiles, while for the hottest star ($T_\mathrm{eff}\approx14000$~K) we used 344 lines. The resulting gain factor ranges from 5 to 17, depending on the $S/N$ of observations and the $T_\mathrm{eff}$ of the star.

For Stokes~$I$, the line weights employed in the LSD analysis are assumed to be the central line depths provided by VALD. For the Stokes~$V$ spectrum, we computed line weights using
\begin{equation}
\label{weightV}
 w = d\,z\frac{\lambda}{\lambda_{0}},
\end{equation}
where $d$ is the central line depth, $z$ is the effective Land\'e factor, $\lambda$ is the laboratory wavelength of the respective line and $\lambda_{0}$ is a normalization wavelength, which we assumed to be equal to 4800~\AA. In a few cases when Land\'e factors of the upper and/or lower atomic levels were missing in VALD, we used the LS-coupling scheme to compute them. In all other cases Land\'e factors come from the quantum-mechanical calculations of Kurucz\footnote{http://kurucz.harvard.edu/linelists.html}. This way of assigning weights is considered standard \citep{Donati:1997}.

\begin{figure}[!th]
   \centering
   \includegraphics[angle=90,width=\columnwidth]{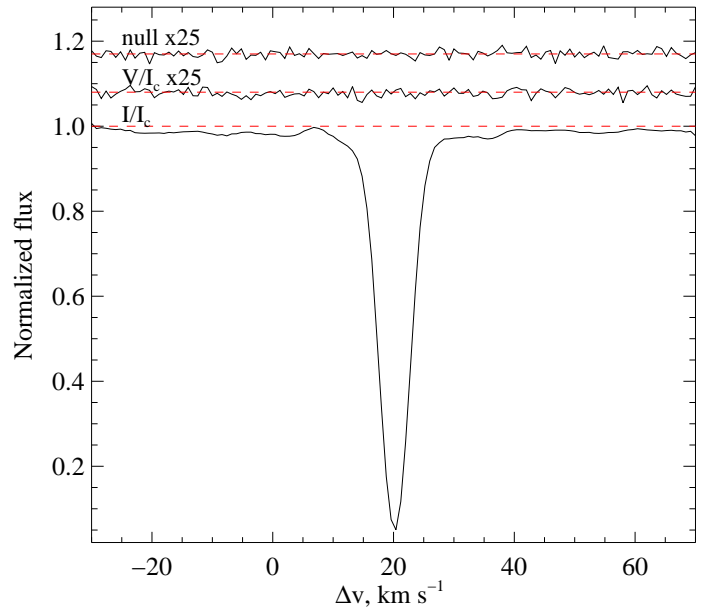}
   \caption{The resulting LSD profiles for HD~71066. For displaying purposes LSD $V$ and null profiles are shifted upwards by 1.08 and 1.17, respectively, and are multiplied by 25.}
   \label{LSDprof}
\end{figure}

We derived the Stokes $I$, $V$ and null LSD profiles using the code written by one of us (Kochukhov et al., submitted). All profiles were reconstructed using a velocity grid with a binsize of 0.8~km\,s$^{-1}$, corresponding to the average pixel scale of HARPSpol spectra. The formal uncertainties of the LSD profiles were obtained by propagating the variance provided by the spectral reduction code for individual pixels. To account for possible under- or over-estimation of the error bars and for the eventual failure of the LSD assumptions, we scaled the uncertainties of the calculated LSD profiles to achieve reduced chi-square values of 1.0, in the same manner as described by \citet{Wade:2000}.
In Fig.~\ref{LSDprof} we show an example of the LSD spectra obtained for the sharp-lined HgMn star HD~71066. This is the object for which we have achieved our lowest uncertainty for the magnetic field diagnostics.

\subsection{Magnetic field measurements}
\label{mf}

The mean longitudinal magnetic field, \Bz, was inferred from LSD profiles. Prior to the magnetic field measurements, we renormalize each LSD profile by a constant factor in order to obtain the correct continuum normalization. The displacement of the continuum is caused by the contribution of numerous unaccounted for weak blends in the LSD~$I$ profile, resulting in a slightly lower continuum level. The scaling of LSD~$I$, $V$ and null profiles was done consistently for each star. Then renormalized LSD profiles were used to estimate the longitudinal magnetic using
\begin{equation}
\label{B}
 \langle B_\mathrm{z} \rangle = -7.14\times10^{6}\frac{\int V(v-v_\mathrm{0})dv}{\int(1-I)dv},
\end{equation}
where $V$ corresponds to the LSD circular polarization spectrum and $I$ corresponds to the LSD intensity profile, $v$ is the velocity, and $v_\mathrm{0}$ is the velocity of the center-of-gravity of the LSD Stokes~$I$ profile. The constant factor in the right hand side of Eq.~(\ref{B}) includes $\lambda_\mathrm{0}$\,--\,the normalization wavelength of the LSD profile, equal to 4800~\AA, and $z_\mathrm{0}$\,--\, Land\'e factor of the LSD profile, equal to 1 according to our definition in Eq.~(\ref{weightV}). The resulting values of \Bz\ are in Gauss. The uncertainties in the \Bz\ measurements are calculated by propagation of the LSD profile uncertainties.

The integration limits in Eq.~(\ref{B}) were chosen on the basis of the LSD~$I$ profile for each star. We note that the choice of integration limits can be quite important for the resulting \Bz\ and its errors bar, especially for stars with $v_\mathrm{e}\sin\,i\la25$~km\,s$^{-1}$. The integration limits must be selected symmetrically with respect to the core of the LSD~$I$ profile. They should not exclude the wings of the profile but, at the same time, should not extend too far into the continuum which contains no useful polarization signal. We find that for very slowly rotating stars the sensitivity of our \Bz\ measurements can be degraded by 5--7~G if the integration limits are changed by $\sim$3~km\,s$^{-1}$. In this case we carefully chose the integration limits in order to reach the best precision of the longitudinal field determination. In the case of faster rotators, the definition of the measurement window is less important, so the integration limits can be set with the precision of $\sim$5~km\,s$^{-1}$.

The same measurement procedure was also applied to the LSD profiles calculated from the null spectrum in order to test whether a real polarization signature is contained in the stellar spectrum or if it is produced by spurious polarization. In general, the uncertainties of the longitudinal magnetic fields determined from the Stokes~$V$ and null spectra are in excellent agreement and in none of the cases do we derive significant longitudinal field from the null profile.  This result confirms the quality of our observations and analysis and suggests the absence of noticeable spurious polarization.

We have measured magnetic fields for 41 objects. Since the target list includes SB2 systems with the contribution of both stars clearly visible in the LSD profiles, we also measured magnetic fields in 6 secondary components. Most of them are Am stars \citep{Ryabchikova:1998}, except HD~27376 where the secondary has Hg peculiarity. There are 68 \Bz\ measurements in total, as some of the stars were observed more than once.

\begin{figure}[t]
   \centering
   \includegraphics[width=\columnwidth,angle=90]{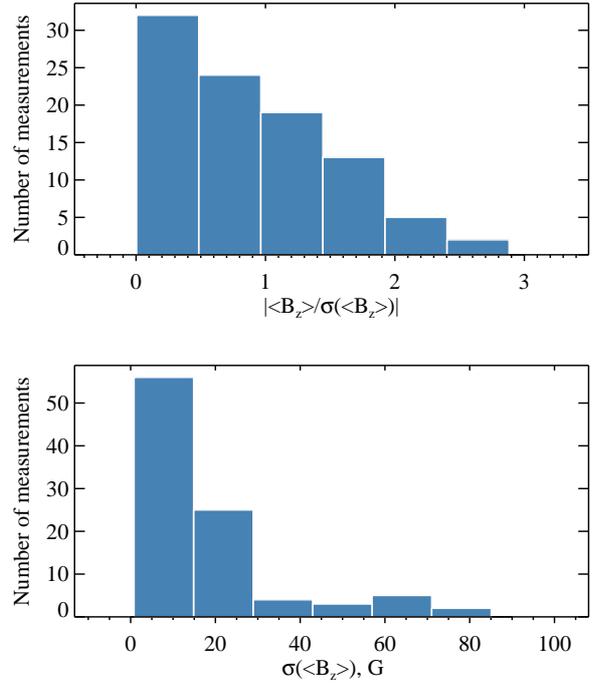}
   \caption{The upper panel: the distribution of the measured longitudinal field divided by its error. The lower panel: the distribution of longitudinal magnetic field error.}
   \label{hist}
\end{figure}

For all stars, our measured \Bz\ values are below the 3$\sigma$ level. Similarly we do not detect \Bz\ using the null spectra. The best precision we achieved for our longitudinal magnetic field measurements was for HD~71066, resulting in an uncertainty of just 0.81~G. Fig.~\ref{hist} presents a statistical summary of our results, excluding measurements of non-HgMn stars. The upper panel shows the distribution of the ratio of \Bz\ to its uncertainty. The lower panel illustrates the overall distribution of the longitudinal field uncertainties. The sensitivity of the majority of our measurements lies in the range of 1--20~G, which is much better than in previous studies of HgMn stars.

\begin{table*}
   \caption{Magnetic field analysis of HgMn stars.}
   \label{tab2}
\centering
\begin{tabular}{l c l l l r r l}
\hline\hline
Target & Comp. & HJD$-24\times 10^{5}$ & $S/N$ & $S/N$(LSD) & \Bz (V), G~ & \Bz (null), G~ & FAP$\times$10 \\
\hline
HD~1909   & A  & 55202.527489 & 200 & 1494 & $-4.02 \pm7.25$    & $9.64 \pm7.11$    & 2.1790 \\
HD~1909   & B  & 55202.527489 & 200 & 1494 & $137.13 \pm58.90$  & $-61.21 \pm57.77$ & 0.4248 \\
HD~1909   & A  & 55210.520641 & 178 & 1262 & $4.60 \pm8.25$     & $-8.42 \pm8.15$   & 1.0770 \\
HD~1909   & B  & 55210.520641 & 178 & 1262 & $47.95 \pm61.14$   & $-99.79 \pm60.43$ & 9.6940 \\
HD~1909   & A  & 55212.522266 & 174 & 1330 & $16.81 \pm8.87$    & $2.77 \pm8.76$    & 4.2570 \\
HD~1909   & B  & 55212.522266 & 174 & 1330 & $12.29 \pm56.73$   & $-48.85 \pm56.08$ & 5.6790 \\
HD~27376  & AB & 55201.768664 & 507 & 3355 & $7.91 \pm3.96$     & $6.82 \pm3.94$    & 4.5860 \\
HD~27376  & A  & 55210.756062 & 422 & 2926 & $-0.30 \pm5.26$    & $10.76 \pm5.23$   & 4.0610 \\
HD~27376  & B  & 55210.756062 & 422 & 2926 & $-8.80 \pm7.01$    & $-5.19 \pm6.98$   & 2.2240 \\
HD~27376  & A  & 55212.741449 & 589 & 4083 & $-2.45 \pm3.89$    & $-2.22 \pm3.88$   & 9.3210 \\
HD~27376  & B  & 55212.741449 & 589 & 4083 & $4.78 \pm4.41$     & $3.55 \pm4.40$    & 9.9540 \\
HD~27376  & A  & 55213.730549 & 411 & 3071 & $5.62 \pm4.91$     & $8.78 \pm4.90$    & 8.4040 \\
HD~27376  & B  & 55213.730549 & 411 & 3071 & $1.13 \pm6.48$     & $-12.67 \pm6.46$  & 3.9460 \\
HD~28217  &    & 55213.538340 & 283 & 1472 & $-12.30 \pm46.87$  & $46.35 \pm46.60$  & 1.4140 \\
HD~29589  &    & 55201.637379 & 267 & 1302 & $-32.34 \pm41.38$  & $-38.16 \pm41.16$ & 0.5699 \\
HD~33647  &    & 55204.752546 & 177 & 1129 & $29.51 \pm38.47$   & $-34.16 \pm37.99$ & 3.8950 \\
HD~33904  &    & 55204.767578 & 600 & 4059 & $-2.55 \pm2.59$    & $1.94 \pm2.57$    & 4.0470 \\
HD~34880  &    & 55201.786748 & 154 & 933  & $55.57 \pm49.14$   & $79.04 \pm47.99$  & 9.2710 \\
HD~35548  &    & 55202.740129 & 172 & 1588 & $-0.74 \pm2.54$    & $0.29 \pm2.51$    & 5.0180 \\
HD~36881  &    & 55200.768514 & 262 & 2451 & $-7.09 \pm5.49$    & $4.58 \pm5.47$    & 8.1070 \\
HD~42657  &    & 55202.757839 & 185 & 1220 & $-30.84 \pm79.97$  & $17.52 \pm79.25$  & 7.4910 \\
HD~53244  &    & 55204.779860 & 477 & 2692 & $14.71 \pm13.06$   & $16.05 \pm12.98$  & 0.0655 \\
HD~53929  &    & 55200.792070 & 319 & 1783 & $7.14 \pm10.77$    & $-13.22 \pm10.59$ & 9.9250 \\
HD~63975  &    & 55204.793412 & 299 & 1828 & $8.28 \pm11.55$    & $-3.83 \pm11.34$  & 6.8330 \\
HD~65950  &    & 55206.754842 & 143 & 1117 & $-0.69 \pm16.87$   & $-24.08 \pm16.35$ & 7.2770 \\
HD~65950  &    & 55211.765235 & 152 & 1039 & $6.78 \pm18.33$    & $12.50 \pm17.77$  & 7.1000 \\
HD~65950  &    & 55212.797482 & 174 & 1355 & $-2.35 \pm13.79$   & $9.05 \pm13.51$   & 3.4760 \\
HD~68099  &    & 55202.773866 & 164 & 1096 & $-108.34 \pm69.86$ & $40.66 \pm68.10$  & 4.9430 \\
HD~68099  &    & 55212.821947 & 206 & 1381 & $96.70 \pm58.42$   & $31.10 \pm57.31$  & 6.1680 \\
HD~70235  &    & 55205.769179 & 178 & 1292 & $22.95 \pm14.90$   & $2.99 \pm14.66$   & 4.9740 \\
HD~70235  &    & 55212.837161 & 204 & 1581 & $30.27 \pm11.93$   & $-14.35 \pm11.72$ & 9.1010 \\
HD~71066  &    & 55205.786739 & 389 & 2730 & $-1.14 \pm0.81$    & $0.24 \pm0.80$    & 0.6773 \\
HD~71833  &    & 55205.812637 & 193 & 1253 & $27.38 \pm26.46$   & $-14.94 \pm25.91$ & 5.7850 \\
HD~71833  &    & 55212.855644 & 153 & 1171 & $-1.69 \pm27.67$   & $20.61 \pm27.06$  & 8.1900 \\
HD~72208  &    & 55202.789124 & 140 & 1289 & $61.39 \pm38.66$   & $-48.43 \pm38.20$ & 7.2000 \\
HD~72208  &    & 55209.769108 & 157 & 1445 & $-2.11 \pm41.08$   & $-46.09 \pm40.41$ & 0.1644 \\
HD~75333  &    & 55206.823570 & 192 & 2155 & $4.38 \pm15.96$    & $-3.40 \pm15.78$  & 3.8920 \\
HD~78316  &    & 55202.803837 & 230 & 1311 & $0.03 \pm4.12$     & $0.42 \pm4.04$    & 3.9410 \\
HD~78316  &    & 55211.787325 & 339 & 1926 & $-0.38 \pm2.66$    & $-3.40 \pm2.64$   & 5.3000 \\
HD~90264  & A  & 55201.850566 & 255 & 1461 & $3.43 \pm8.39$     & $-33.70 \pm8.23$  & 5.7250 \\
HD~90264  & B  & 55201.850566 & 255 & 1461 & $3.34 \pm12.42$    & $-2.50 \pm12.17$  & 0.6511 \\
HD~101189 &    & 54982.974439 & 51  & 262  & $-95.22 \pm65.63$  & $-53.75 \pm62.58$ & 0.4069 \\
HD~101189 &    & 55201.862940 & 179 & 2797 & $-6.41 \pm5.43$    & $4.52 \pm5.37$    & 1.4030 \\
HD~106625 &    & 54983.067376 & 149 & 2493 & $-15.99 \pm18.20$  & $13.06 \pm18.41$  & 4.1200 \\
HD~106625 &    & 55205.872226 & 881 & 4348 & $14.71 \pm11.11$   & $7.98 \pm11.13$   & 10.000 \\
HD~110073 &    & 55205.856731 & 417 & 2813 & $2.96 \pm4.25$     & $1.71 \pm4.26$    & 9.9040 \\
HD~141556 & A  & 54982.163224 & 120 & 1613 & $-1.91 \pm3.46$    &                   & 7.4680 \\
HD~141556 & B  & 54982.163224 & 120 & 1613 & $8.38 \pm10.81$    &                   & 1.4170 \\
HD~141556 & A  & 55317.770736 & 143 & 1710 & $1.54 \pm4.48$     & $-12.63 \pm4.51$  & 2.9260 \\
HD~141556 & B  & 55317.770736 & 143 & 1710 & $17.90 \pm11.08$   & $8.03 \pm11.16$   & 1.9210 \\
HD~149121 &    & 55316.735917 & 217 & 1840 & $-0.02 \pm4.45$    & $1.04 \pm4.48$    & 6.4580 \\
HD~158704 &    & 55316.845983 & 172 & 967  & $-0.76 \pm3.56$    & $-1.91 \pm3.50$   & 2.6810 \\
HD~165493 & A  & 55316.925608 & 144 & 796  & $4.77 \pm5.56$     & $-2.10 \pm5.51$   & 8.1200 \\
HD~165493 & B  & 55316.925608 & 144 & 796  & $-31.05 \pm27.67$  & $-80.86 \pm27.42$ & 0.9936 \\
HD~175640 &    & 55317.865578 & 170 & 1327 & $-0.59 \pm2.16$    & $-3.05 \pm2.16$   & 8.4280 \\
HD~178065 &    & 55317.879714 & 57  & 431  & $-1.92 \pm8.81$    &                   & 9.6770 \\
\textbf{HD~178065} &    & \textbf{55319.831679} & \textbf{147} & \textbf{1323} & \textbf{$-2.26 \pm2.08$}    & \textbf{$-4.12 \pm2.08$}   & \textbf{0.0085} \\
HD~179761 &    & 55319.926735 & 123 & 1048 & $23.92 \pm19.26$   & $-21.95 \pm19.07$ & 5.1960 \\
HD~186122 &    & 55319.850529 & 203 & 1252 & $0.66 \pm1.85$     & $0.45 \pm1.83$    & 7.3990 \\
HD~191110 & A  & 55319.934190 & 143 & 1256 & $1.67 \pm7.17$     & $4.27 \pm7.13$    & 9.8480 \\
HD~191110 & B  & 55319.934190 & 143 & 1256 & $6.26 \pm9.15$     & $20.55 \pm9.10$   & 7.2040 \\
HD~193452 &    & 55316.939679 & 90  & 964  & $-2.10 \pm2.37$    &                   & 2.2220 \\
HD~193452 &    & 55319.868330 & 203 & 1868 & $-1.04 \pm1.42$    & $-0.70 \pm1.41$   & 2.0460 \\
HD~194783 &    & 54984.354843 & 160 & 1316 & $10.08 \pm19.75$   & $-3.23 \pm20.05$  & 8.1140 \\
HD~194783 &    & 55319.945379 & 136 & 1064 & $44.01 \pm21.60$   & $24.02 \pm21.07$  & 4.0580 \\
HD~202671 &    & 54984.382173 & 175 & 1220 & $25.51 \pm16.01$   & $-25.08 \pm16.23$ & 2.3420 \\
HD~211838 &    & 54984.396740 & 150 & 1432 & $-194.23 \pm77.70$ & $-2.01 \pm79.02$  & 7.1810 \\
HD~221507 &    & 55210.548529 & 349 & 2498 & $4.21 \pm7.04$     & $-9.51 \pm6.99$   & 9.4710 \\
\hline
\end{tabular}
\end{table*}

For certain magnetic field configurations the first moment of the LSD Stokes~$V$ profile appearing in Eq.~(\ref{B}) may be zero, while the profile itself shows a magnetic signature. In order to examine such situations we employed the False Alarm Probability (FAP) analysis \citep{Donati:1997}. It is a $\chi^2$ probability statistic, which estimates the probability that a given signal is produced by random noise. We employed FAP analysis as a tool for the detection of a magnetic signal in the LSD Stokes $V$ spectra, assuming \textit{no detection} if FAP~$\,>\,$10$^{-3}$, \textit{marginal detection} if 10$^{-3}\le\,$FAP$\le\,$10$^{-5}$ and \textit{definite detection} if FAP$\,<\,$10$^{-5}$.
We report no detection of any signatures indicative of a magnetic field in any of the 41 HgMn stars we observed. There is also no evidence of the coherent signal in any of the null spectra. For one star, HD~178065, the FAP analysis of the LSD Stokes~$V$ yielded a marginal detection. This result could be a statistical fluctuation since the measured FAP is close to the (arbitrary) boundary between \textit{no detection} and \textit{marginal detection}. No longitudinal magnetic field was detected for this star (\Bz$=-2.26 \pm2.08$~G).

A complete summary of our analysis of the circular polarization spectra of HgMn stars is presented in Table~\ref{tab2}. In the columns from left to right we list the stellar HD number, designation of a binary component (A -- for primary, B -- for secondary and AB -- for one measurement of the blended components of HD~27376). This is followed by columns containing the Julian date of observation, the signal-to-noise ratio of the observed Stokes~$V$ spectrum and of the reconstructed LSD Stokes~$V$ profile, the mean longitudinal magnetic field derived from the Stokes~$V$ and null spectra, together with the respective uncertainties. In a few cases polarization measurements were based on 2 sub-exposures only, which did not allow us to compute the null spectrum. The uncertainties correspond to 1$\sigma$ confidence levels. The last column gives the FAP for the presence of a signal in the LSD Stokes~$V$ profile. All LSD Stokes~$I$, $V$ and null spectra are presented in Fig.~\ref{LSD_all} (Online material).

\section{Discussion}
\label{disc}

As already mentioned, previous studies of magnetic fields in HgMn stars did not detect longitudinal fields stronger than 29--100~G, and in one particular case the limits were 6--19~G. These 1~$\sigma$ error bars are not low enough to exclude the presence of the longitudinal fields weaker than $\sim$100~G, which could result from dynamically important local magnetic field structures covering only a fraction of the stellar surface. Thus, our understanding of the magnetic characteristics of these stars was fundamentally incomplete. In order to achieve a more detailed picture of the magnetic properties of HgMn stars, we studied a large sample of such stars with the most stable and modern spectropolarimeter available.

Our analysis of the Stokes~$V$ spectra showed no evidence of any polarimetric signal, setting 3$\sigma$ upper limits on the longitudinal field present of 3--30~G for the majority of the stars we observed. For one sharp-lined HgMn star we obtained an uncertainty of $\sigma_{B_z}<$1~G, which is better than previously achieved with a single observation of any early-type star. In addition to obtaining null results for longitudinal magnetic field measurements, we analysed the Stokes~$V$ LSD profiles themselves, finding no evidence for complex polarization signatures. Thus, we conclude that mercury-manganese stars do not possess strong, but topologically complex, magnetic fields.
If we restrict ourselves to a dipolar magnetic field topologies, the dipole strength can be roughly estimated as $B_\mathrm{d}\sim3$~\Bz. Based on our longitudinal field measurements, we conclude that the HgMn stars do not possess dipolar fields stronger than 3--30~G.

It is important to evaluate our results in the context of the possible role of magnetic fields in the recently discovered spot formation on HgMn stars. Although there are no detailed models describing the process of the interaction between magnetic fields and the plasma, it is understood \citep[e.g.,][]{Wade:2006,Auriere:2007} that this interaction becomes important when the strength of magnetic field exceeds the equipartition limit, defined by \citet{Wade:2006} as $B_\mathrm{eq}=\sqrt{12\pi P_\mathrm{gas}}$. At this field strength the energy of the magnetic field equals that of the gas.

\begin{figure}[t]
   \centering
   \includegraphics[angle=90,width=\columnwidth]{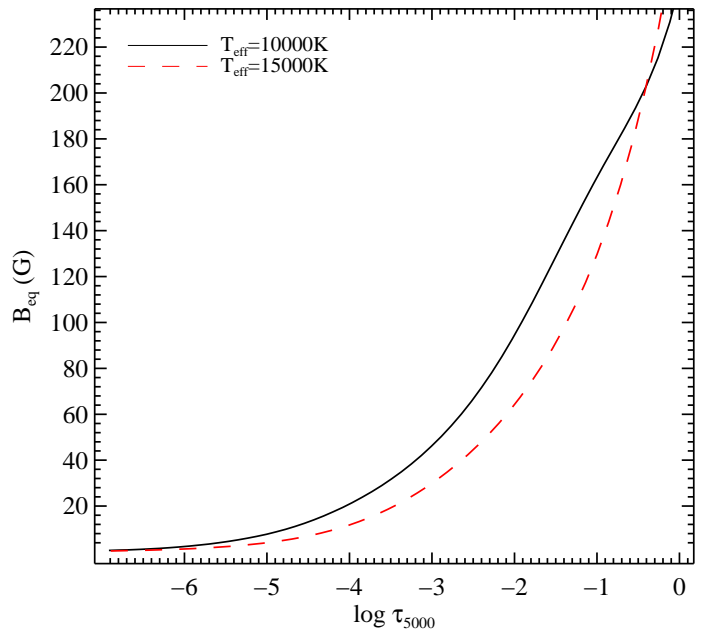}
   \caption{The equipartition magnetic field as a function of the optical depth. The solid line corresponds to the cooler model and the dashed one to the hotter.}
   \label{Beq}
\end{figure}

We estimated the value of the equipartition magnetic field strength for atmosphere models with effective temperatures of $T_\mathrm{eff}$=10000~K and $T_\mathrm{eff}$=15000~K, representing the range covered by our sample of HgMn stars. Using thermodynamic parameters tabulated in these models and assuming the ideal gas law, we calculated the equipartition field as a function of the optical depth in the continuum. The result is presented in Fig.~\ref{Beq}. From this figure we can see that the value of $B_\mathrm{eq}$ at $\log\tau_{5000}=-0.5$ is $\sim$200~G for the stars of both low and high effective temperatures.  The value of $B_\mathrm{eq}$ then decreases to 100~G at $\log\tau_{5000}=-2$ for $T_\mathrm{eff}=$10000~K and to 60~G for $T_\mathrm{eff}=$15000~K. The majority of spectral lines used in our measurements form over a range in optical depth from $\log\tau_{5000}=-0.5$ to $-2$. In this interval the values of $B_\mathrm{eq}$ are much higher than our upper limits for the organized magnetic field on the stellar surface. Thus, it appears there are no magnetic fields on the surfaces of HgMn stars that could lead to the formation of chemical spots. These conclusions apply to the fields stronger than $\sim$\,10~G probed by our study. On the other hand, we cannot exclude the presence of complex magnetic fields with a field strength of $\sim$\,1~G in surface spots, similar to the tangled magnetic topology suggested for Vega \citep{Petit:2010}.

These results lead to the conclusion that there exists a real dichotomy between HgMn and magnetic Ap stars. The latter show chemical inhomogeneities in the presence of dipolar-like magnetic fields stronger than 300~G. This limit is determined by the field strength that can withstand shearing by differential rotation and survive at the surfaces of early-type stars \citep{Auriere:2007}. In contrast to magnetic Ap stars and despite a similarity in fundamental parameters, HgMn stars clearly do not have magnetic fields reaching this limit. Consequently, the surface chemical structures in mercury-manganese stars must form under a mechanism fundamentally different from magnetically-driven spot formation in hot Ap or cool active stars.

A detailed Doppler imaging study of the surface structures on the brightest HgMn star, $\alpha$~And, carried out by \citet{Kochukhov:2007} using observations collected over a period of seven years led to the remarkable discovery of Hg spot \textit{evolution}. This is very different from the spotted early-type magnetic stars, in which the morphology of spots remains steady over tens of years. The fact that no magnetic field was detected in $\alpha$~And \citep{Wade:2006} makes the explanation of the spot formation and evolution a very challenging task. \citet{Kochukhov:2007} suggested that mercury atoms are concentrated in a thin layer where the radiation pressure and gravitational force compensate for each other. The observed changes in the spot morphology are then ascribed to hydrodynamical instabilities, upsetting the fine force balance in the Hg cloud. The physics of this intricate process and its relation to the binarity, stellar rotation and dynamic atomic diffusion effects is yet to be explored.

Despite being frequent targets of detailed abundance analyses during more than half a century, HgMn stars were recognized as spotted variables only recently. To date, 7 spotted HgMn stars have been found. It is uncertain how many more such stars will be discovered. Prior to the formulation of exotic theories of non-magnetic chemical spot formation, it is important to determine a complete picture of the occurrence rate of chemical inhomogeneities on the surface of HgMn stars. If spots are found in only a small fraction of these objects, it may be that the observed structure formation is associated with a certain stage of stellar evolution or a certain configuration of the binary system. On the other hand, discovery of spots in the majority of HgMn stars would indicate the presence of a universal hydrodynamical process, not anticipated in current stellar models.

We believe that in-depth studies of selected spotted HgMn stars as well as a broad search for spectral line variability in all accessible objects of this type will provide sufficient observational constraints for the development of a theory of non-magnetic spot formation. Analysis of known spotted stars can yield information on the time-scales of spot evolution and, possibly, on the heights of these inhomogeneities relative to the typical line formation regions. It is also important to continue searching for weak magnetic fields in spotted HgMn stars as previously done by \citet{Wade:2006} for $\alpha$~And and by \citet{Folsom:2010} for AR~Aur. On the other hand, a line profile variability survey of a large sample of HgMn stars will help us to uncover dependencies on stellar parameters, rotation rate or binarity, possibly revealing a correlation of one these factors with the presence of spots.  In subsequent studies of HgMn stars we plan to carry out both detailed studies of individual objects and to perform variability surveys to find new spectrum variable stars.

\begin{acknowledgements}
Authors thank the referee Pascal Petit for his valuable comments. VM and NP thank the European Southern Observatory (ESO) engineering team for excellent support during the commissioning of HARPSpol. OK is a Royal Swedish Academy of Sciences Research Fellow, supported by grants from Knut and Alice Wallenberg Foundation and Swedish Research Council.
\end{acknowledgements}

\Online
\onlfig{5}{
\begin{figure*}
 \includegraphics[width=17.5cm]{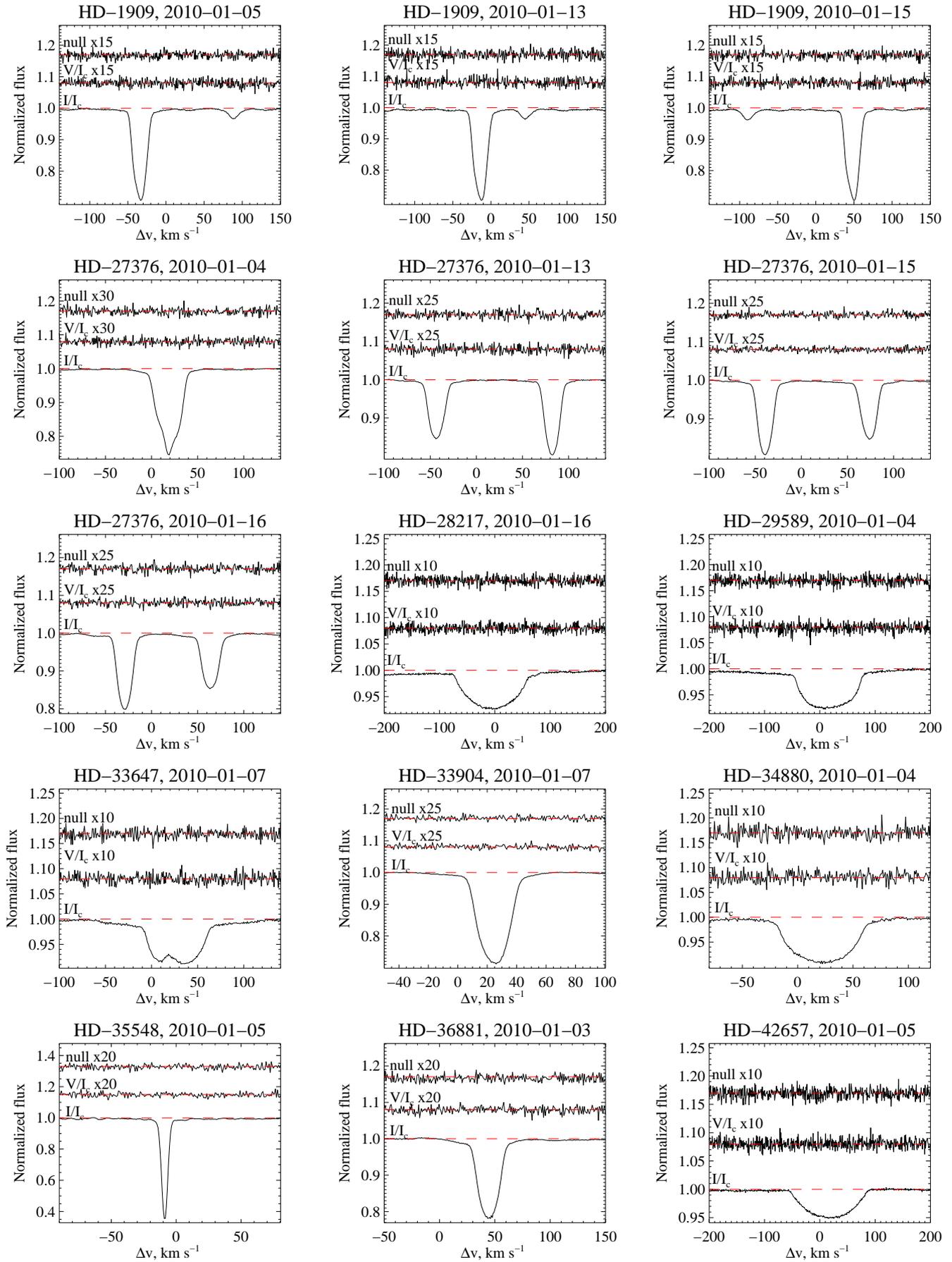}
 \caption{From bottom to top each panel shows the LSD profiles of Stokes~I, V and null spectrum (if available) for all HgMn stars in our survey.}
 \label{LSD_all}
\end{figure*}}

\onlfig{5}{
\begin{figure*}
 \includegraphics[width=17.5cm]{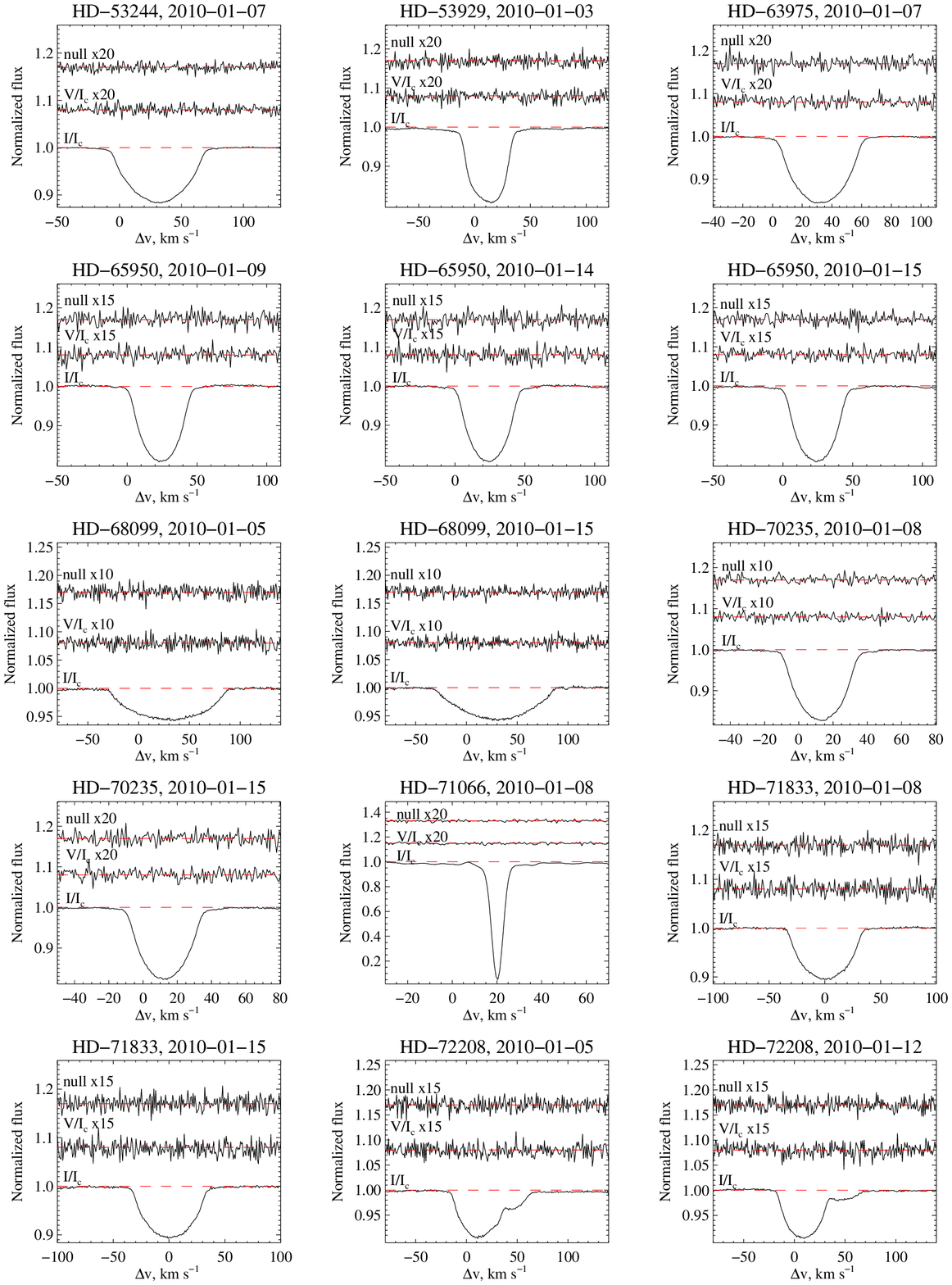}
 \caption{Continued.}
\end{figure*}}

\onlfig{5}{
\begin{figure*}
 \includegraphics[width=17.5cm]{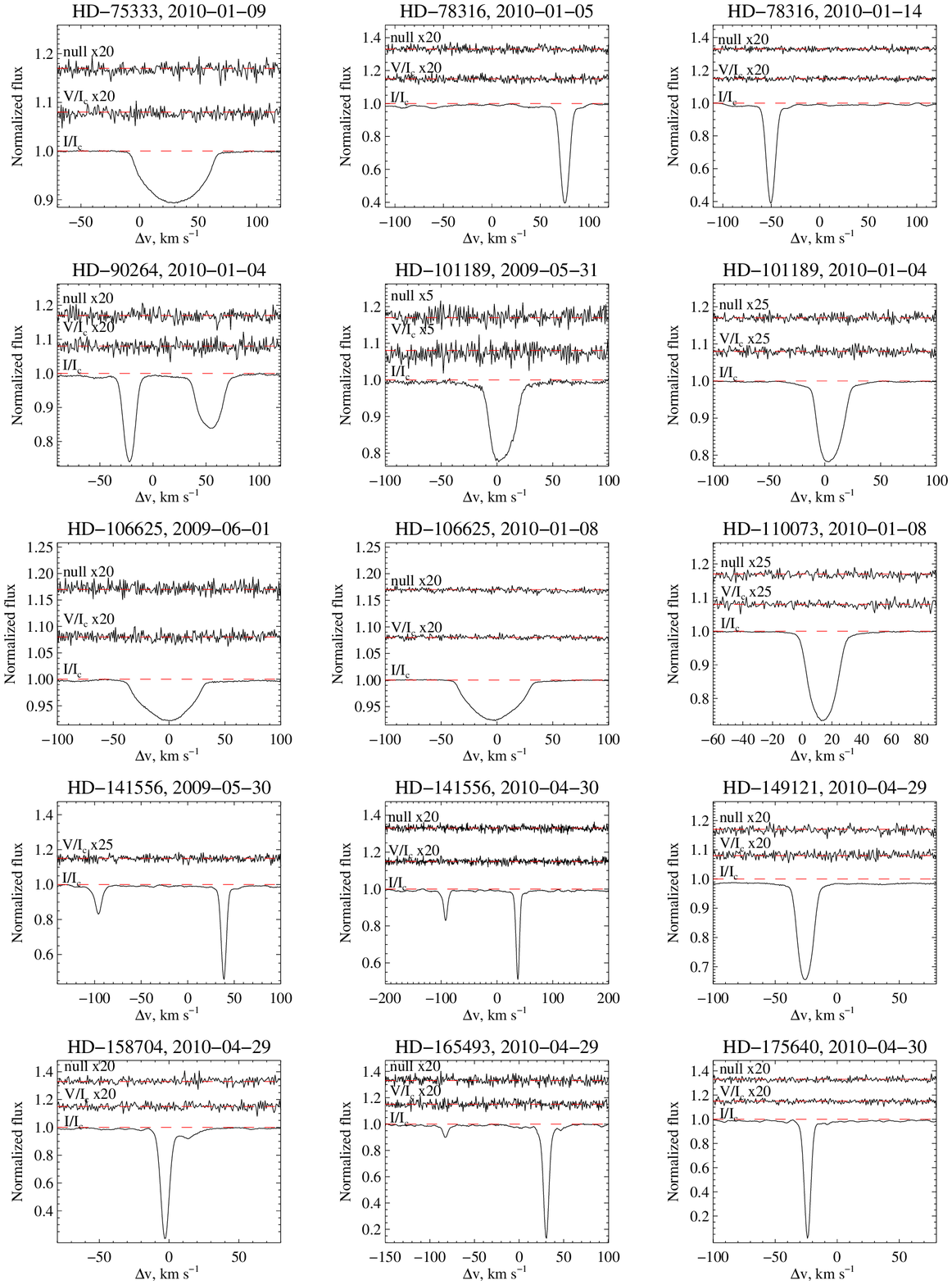}
 \caption{Continued.}
\end{figure*}}

\onlfig{5}{
\begin{figure*}
 \includegraphics[width=17.5cm]{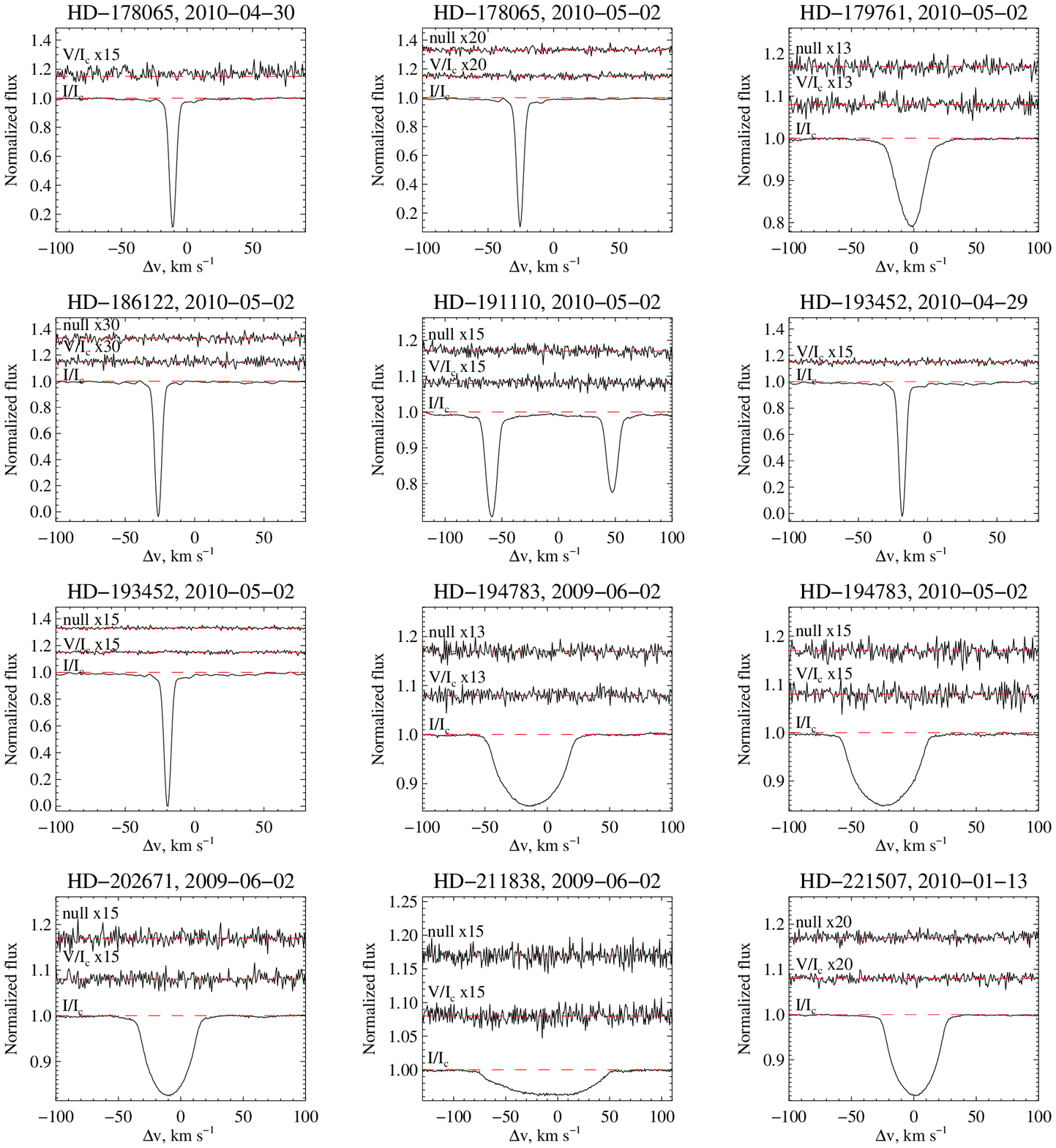}
 \caption{Continued.}
\end{figure*}}

\end{document}